\documentclass[draft]{agujournal2019}
\usepackage{url} 
\usepackage{lineno}
\usepackage{amsmath}
\usepackage[inline]{trackchanges} 
\usepackage{soul}
\usepackage{comment}
\usepackage{natbib}
\usepackage{float} 
\draftfalse

\journalname{JGR: Machine Learning and Computation}

\graphicspath{{Figures/}}
\begin{document}

%
%


\title{An effective physics-informed neural operator framework for predicting wavefields}

%
%




\authors{
X. Ma\affil{1}, 
T. Alkhalifah\affil{1}
}


\affiliation{1}{Earth Science and Engineering Program, Physical Science and Engineering Division, King Abdullah University of Science and Technology (KAUST),Thuwal, 23955-6900, Kingdom of Saudi Arabia}





\correspondingauthor{Xiao Ma}{xiao.ma@kaust.edu.sa}



\begin{keypoints}
\item We propose a physics-informed convolutional neural operator (PICNO) to efficiently predict wavefields governed by the Helmholtz equation.
\item By embedding physics-based constraints into the training process, PICNO significantly improves generalization over conventional data-driven neural operators.
\item PICNO achieves up to 53\% reduction in relative error and more physically consistent predictions, even under high-frequency conditions or out-of-distribution scenarios.
\end{keypoints}

%
%

%
%


\begin{abstract}
Solving the wave equation is fundamental for geophysical applications. However, numerical solutions of the Helmholtz equation face significant computational and memory challenges. Therefore, we introduce a physics-informed convolutional neural operator (PICNO) to solve the Helmholtz equation efficiently. The PICNO takes both the background wavefield corresponding to a homogeneous medium and the velocity model as input function space,  generating the scattered wavefield as the output function space. Our workflow integrates PDE constraints directly into the training process, enabling the neural operator to not only fit the available data but also capture the underlying physics governing wave phenomena. PICNO allows for high-resolution reasonably accurate predictions even with limited training samples, and it demonstrates significant improvements over a purely data-driven convolutional neural operator (CNO), particularly in predicting high-frequency wavefields. These features and improvements are important for waveform inversion down the road.

\end{abstract}

\section*{Plain Language Summary}
Seismic wavefields are essential for imaging the Earth’s subsurface, but traditional numerical methods to simulate them can be very slow, especially when dealing with complex geological models or high-frequency waves. In this study, we introduce a machine learning approach that not only learns from existing data, but also follows the laws of physics governing how waves travel underground. Our method uses a type of neural network called a convolutional neural operator, enhanced with physical rules, to predict wavefields much faster than conventional solvers. This physics-informed approach improves both accuracy and reliability, even when applied to unseen geological settings. It offers a promising path to accelerate seismic simulations for large-scale applications like oil exploration or earthquake research.

%
%

%


%
%
%
%

\section{Introduction}
Seismic imaging and inversion have long been important topics in the geophysical exploration and seismology community, as they play a critical role in delineating subsurface structures and characterizing geophysical properties. At the heart of these techniques lies wavefield forward modeling, which serves as the physical foundation for reconstructing the Earth's interior. Accurately computing seismic wavefields is essential for improving the fidelity of inversion results, particularly in complex geological settings. Frequency-domain wave equations offer an efficient framework for simulating steady-state responses of seismic waves, and their numerical solutions are traditionally obtained using well-established methods such as finite-difference \citep{berkhout1981wave, wu2018highly, yang2024frequency}, spectral element \citep{berkhout1981wave, gajewski1987computation}, or pseudo-spectral schemes \citep{carcione2010generalization,virieux2011review,wu2018efficient}. These methods have demonstrated significant success across both exploration geophysics and seismological applications. However, their computational costs scale poorly with increasing model size and frequency content, making them increasingly impractical for large-scale 3D simulations or high-resolution imaging tasks. Furthermore, repeated wavefield evaluations for multiple sources and iterations in full waveform inversion (FWI) exacerbate the computational burden, prompting the need for more scalable and flexible modeling approaches. 

Recently, the term "scientific machine learning" (SciML) has emerged as a discipline within the data science and deep learning community. SciML seeks to address domain-specific data challenges and extract insights from physical data through innovative ML-based methodological solutions. SciML draws on tools from both machine learning and scientific computing to develop new methods that are more physics-aware and robust with data analysis. It will be critical in driving the next wave of data-driven scientific discovery in the physical and engineering sciences. Among many SciML methods, Physics-Informed Neural Networks (PINNs) offer a framework for leveraging known physical laws to accurately model the behavior of various physical systems \citep{cuomo2022scientific, huang2022pinnup, song2022versatile}. PINNs aim to approximate the solution of a given system, typically defined by a partial differential equation (PDE) along with initial and boundary conditions (ICs and BCs), by minimizing a physics-constrained loss function \citep{raissi2019physics}. Through the use of automatic differentiation provided by deep learning libraries (pytorch, tensorflow), PINNs compute the derivatives required to evaluate the residuals of the governing PDE. In the geophysics community, PINNs already have shown promising results, including wavefield prediction \citep{song2021solving, alkhalifah2021wavefield,rasht2022physics}, traveltime reconstruction \citep{bin2021pinneik,taufik2023neural,grubas2023neural}, velocity inversion \citep{song2021solving,xue2025multi}. While PINNs have demonstrated considerable success in modeling individual physical scenarios, their limited ability to generalize across varying initial conditions significantly restricts their applicability as efficient surrogate models. 

Another important SciML-based framework is the neural operator, which aims to address the limitations of PINNs. Neural operators are neural networks that learn the mapping between function spaces, offering an appealing alternative by providing fast surrogate models for parametric partial differential equations (PDEs). Different from PINNs, neural operators are a generalized framework for neural networks aimed at learning operators that map between infinite-dimensional function spaces rather than single physical systems (instance), making it a suitable surrogate model for real word modeling. In the architecture of neural operators, a composition of linear integral operators and nonlinear activation functions enables the resulting neural operator to capture complex, nonlinear mappings between the input and the desired output. These neural operators exhibit resolution invariance, offering significant efficiency advantages over conventional neural networks. There are various types of neural operators that have been proposed in recent works, such as Graph Neural Operator (GNO) \citep{li2020neural}, DeepONets \citep{lu2021learning}, Low-rank Neural Operator (LNO) \citep{ryu2024operator}. In particular, the Fourier Neural Operator (FNO) has demonstrated state-of-the-art performance in many applications, such as in weather forecasting \citep{pathak2022fourcastnet} and carbon dioxide geological storage \citep{wen2023real}. In their application, FNO has shown very promising results regarding the computational efficiency and accuracy \citep{li2020fourier}.

In the context of learning geophysical simulations, \cite{yang2023rapid} demonstrated that neural operator-based time-domain full waveform modeling can be nearly two orders of magnitude faster than traditional numerical methods. By training the Fourier Neural Operator (FNO) on a dataset of 20000 velocity models, given by Gaussian random fields (GRF), their approach achieves accurate simulations across a variety of velocity perturbations, source locations, and mesh discretizations. Also, in frequency domain wavefield prediction, \cite{zou2024deep} utilized a U-shape neural operator (U-NO) to learn the Helmholtz operator efficiently, U-NO could predict a relatively accurate frequency domain wavefield compared with the numerical solution in a very short time. In addition, to mitigate the generalization problem, \cite{wang2024transfer} proposed a transfer learning method in which the baseline FNO model is trained at a single source location and frequency and then shared with the target FNO for seamlessly predicting the frequency-domain wavefields at different sources and frequencies. However, this method requires massive training data, which is not available in real situations \citep{song2024seismic}. To allow for a more seamless integration of the source and frequency information into an FNO frequency-domain wavefield predictor, \cite{huang2025learned} used the background, analytically computed, wavefield as input to the neural operator. However, due to the purely data-driven nature of the training, the method also required massive training data or otherwise lacked generalization capabilities. 

Thus, despite the demonstrated success of neural operators on two-dimensional PDE problems, ensuring robust generalization remains an ongoing challenge. Neural operators trained on limited or specific training data distributions may struggle to extrapolate effectively to other data distributions, limiting their applicability in real-world scenarios. The study of generalization for neural operators is still in its early stages. One possible solution for the generalization problem is the use of a physics-informed neural operator (PINO). Recently, researchers extended the "physics-informed" idea into the training of neural operators by embedding the discrete form of PDEs into the loss function \citep{li2024physics}. More specially, by incorporating the violation of physical laws into the loss function, the physics-informed neural operators are encouraged to learn both the data and the underlying governing equations. Instead of relying on automatic differentiation—which is often memory-intensive for this type of architecture—these networks employ the Fourier-based derivative or finite-difference method to calculate the PDE constraints more efficiently. \cite{karniadakis2021physics} showed that PINO generates more accurate simulation results by bridging the gap between physics-informed optimization and data-driven neural operator learning. Several examples show that PINO has a better ability for generalization. Furthermore, \cite{rosofsky2023applications} extended the application of PINO to coupled PDEs. After multiple experiments, they concluded that PINO provides the flexibility to produce different initial and boundary conditions for operator learning. In contrast to traditional neural operators, PINOs excel at enforcing local physics constraints and do not require extensive runs of conventional solvers when preparing the training data. 

In this study, we propose a physics-informed convolutional neural operator (PICNO) to tackle the generalization problem of learning the Helmholtz equation. Our PICNO can achieve satisfactory wavefield predictions even with limited training data. Furthermore, PICNO demonstrates strong performance when applied to complex velocity models and high-frequency wavefield cases. First, we briefly describe the convolutional neural operator (CNO) architecture and corresponding physics-informed loss function. Then, we demonstrate the performance of our approach on the synthetic datasets for different frequencies. Finally, we discuss several existing challenges associated with physics-informed neural operators.
\vspace{0.4cm}

\section{Theory}
\subsection{Problem Statement}

In this study, we aim to develop a neural operator-based framework to model (learn) seismic wavefields in the frequency domain by leveraging the Convolutional Neural Operator. Specifically, our objective is to learn the mapping from velocity models and physical conditions (source location, frequency) to the complex-valued wavefield solution governed by the Helmholtz equation, which describes steady-state wave propagation in heterogeneous media. Let $U(x,z,\omega)$ denote the full complex-valued wavefield at two-dimensional spatial coordinates $(x,z)$ for angular frequency $\omega$. The governing equation is expressed as:
\begin{equation}
L(v, \omega)\, U \;=\; f(x,z),
\label{ho_eq}
\end{equation}
where $v$ is the spatially varying velocity model, and $f(x,z)$ represents the external source term. $L$ is the linear wave equation operator. In the frequency domain, we often refer to Equation~\ref{ho_eq} as the Helmholtz equation.

In practice, numerically modeling the full wavefield $U$ poses challenges due to the entanglement of source and velocity effects \citep{alkhalifah2021wavefield}. To better decouple these influences and focus learning on velocity-induced scattering, we reformulate the learning target for an ML-based modeling as the scattered wavefield $\delta U$, which is defined as the difference between the full wavefield and the background wavefield:
\begin{equation}
\left\{
\begin{aligned}
    &\frac{\omega^2}{v^2} \delta U + \nabla^2 \delta U + \omega^2 \left( \frac{1}{v^2} - \frac{1}{v_0^2} \right) U_0 = 0, \\
    &\delta U = U - U_0.
\end{aligned}
\right.
\end{equation}
Here, $v_0$ is a reference (or background) velocity model, and $U_0$ denotes the corresponding background wavefield, which we simplify by making the background velocity constant. In this case, we can compute $U_0$ analytically using the Green’s function for a constant velocity model, as follows:
\begin{equation}
U_{0}(x,z) = \frac{i}{4} H_{0}^{(2)}
\Biggl(
\omega \sqrt{
\frac{(x - x_s)^2 + (z - z_s)^2}{v_{0}^2}
}
\Biggr),
\end{equation}
where $H_{0}^{(2)}$ denotes the Hankel function of the second kind of order zero, and $(x_s, z_s)$ is the source location. This analytical formulation incorporates both the source position and frequency into the wavefield representation, thereby implicitly encoding key physical information without the need for explicit source embedding within the network.

Following the findings of \cite{huang2024learned}, incorporating $U_0$ as an input feature not only improves convergence, but also provides the necessary physical context for wavefield prediction. Therefore, in this study, both the CNO and the physics-informed CNO (PICNO) are trained to learn the mapping:
\[
(U_0^{\mathrm{Re}},\ U_0^{\mathrm{Im}},\ v) \longrightarrow (\delta U^{\mathrm{Re}},\ \delta U^{\mathrm{Im}}),
\]
where the input channels consist of the real and imaginary parts of the background wavefield and velocity models, and the output channels represent the real and imaginary parts of the scattered wavefield. This formulation enables the network to learn an approximation (operator) that maps physical input fields to seismic responses in the frequency domain, which is essential for efficient forward modeling and inversion in large-scale geophysical applications.

\subsection{The Architecture of CNO}

Convolutional Neural Operators (CNOs) are designed to bridge the gap between continuous mathematical operators and discrete learning architectures. In contrast to traditional neural networks that operate on fixed-dimensional vectors, operator learning frameworks such as CNO aim to learn mappings between infinite-dimensional function spaces. Given their locality, weight-sharing, and computational efficiency, convolutional neural networks (CNNs) serve as a natural foundation for constructing such operators. Moreover, CNOs are constructed to respect the \emph{continuous-discrete equivalence} (CDE), ensuring that the discrete computations performed by the network are faithful approximations of their continuous analogs. The CNO is defined as a compositional mapping between functions:
\begin{equation}
G : u \mapsto P(u)
= v_0 \mapsto v_1 \mapsto \dots \mapsto v_L
\mapsto Q(v_L)
= \overline{u},
\end{equation}
where $P$ is a \emph{lifting operator} that projects the low-dimensional input function $u$ to a high-dimensional latent representation $v_0$, and $Q$ is a \emph{projection operator} that maps the final latent output $v_L$ back to the original function space to obtain the predicted output $\overline{u}$. Each intermediate representation $v_l$ is transformed through a layer-wise operation defined as:
\begin{equation}
v_{l+1} = V_{l} \circ \Sigma_{l} \circ K_{l}(v_{l}), 
\quad 0 \leq l \leq L-1.
\end{equation}

In this formulation, each layer consists of three elementary mappings:
\begin{itemize}
    \item $K_l$: a convolution operator that applies localized filters across the domain;
    \item $\Sigma_l$: a non-linear activation function, for which we use Leaky ReLU (LReLU);
    \item $V_l$: a resolution-changing operator, either upsampling or downsampling, depending on the network stage.
\end{itemize}

\begin{figure}[!t]
\centering
\includegraphics[width=\textwidth]{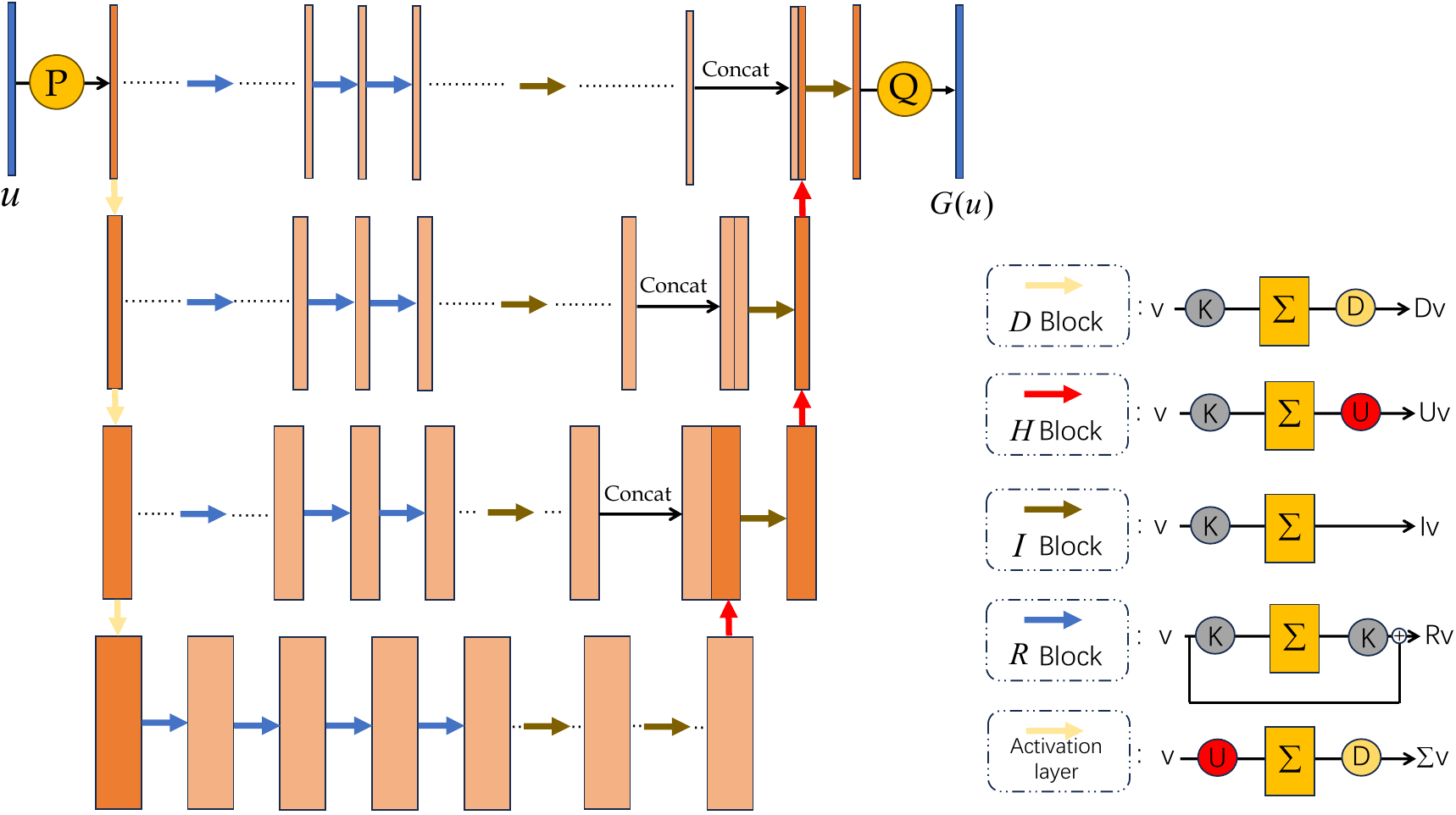}
\caption{ The architecture of CNO. Different colors of the arrows represent different blocks. The input function is first transformed into a high-dimensional space through a lifting layer ($P$), then processed by a series of neural operator layers, and finally mapped back to its original spatial dimensions via a projection operator ($Q$). }
\label{fig1} 
\end{figure}

The convolution operator $K$ is defined in discrete physical space using kernels of size $k \times k$:
\begin{equation}
K(i, j) = \sum_{m=0}^{k-1} \sum_{n=0}^{k-1} w(m, n) \cdot v(i + m, j + n)
\label{eq:conv2d}
\end{equation}

where $w(m, n)$ denotes the trainable weight. Unlike spectral neural operators (e.g., FNO), which represent convolution in the frequency domain via Fourier transforms, CNO performs convolution directly in the spatial domain. This physical-space parameterization provides improved locality and interpretability, and ensures compatibility with non-periodic boundary conditions and arbitrary geometries. The upsampling operator $H$ increases resolution by inserting zero-valued samples between adjacent points, followed by a low-pass filter to suppress high-frequency artifacts introduced during interpolation. Conversely, the downsampling operator $D$ first applies a low-pass filter to the input, and then reduces its resolution through subsampling, thereby mitigating aliasing effects.To enable expressive feature representations and support multiscale learning, CNO adopts a modular architecture composed of four key building blocks. Each of these blocks is constructed through a composition of the elementary mappings introduced earlier—namely, convolution ($K$), nonlinearity ($\Sigma$), and resolution adjustment ($V$):
\begin{enumerate}
    \item \textbf{Downsampling block ($D$)}: Applies the downsampling operator to reduce spatial resolution, enabling the extraction of coarse-grained features.
    
    \item \textbf{Upsampling block ($H$)}: Reconstructs higher-resolution features from coarser representations through interpolation and filtering.
    
    \item \textbf{Invariant block ($I$)}: Combines a standard convolution with a non-linear activation function to maintain spatial resolution while refining features.
    
    \item \textbf{Residual block ($R$)}: Employs skip connections to enhance gradient flow and stabilize training, thereby improving the model’s capacity to capture complex mappings.

\end{enumerate}
Each block receives a band-limited function as input and produces another band-limited function of equivalent frequency content, thereby preserving the signal structure throughout the network. To enhance feature propagation and promote the reuse of intermediate representations, the U-Net-style patching is employed, whereby outputs from earlier layers are concatenated with those from later layers as additional channels.
Figure~\ref{fig1} illustrates the overall architecture of CNO. The input function $u$ is first transformed by the lifting operator $P$ into a high-dimensional latent space. This representation is then passed through a sequence of neural operator layers, each composed of convolution, activation, and resolution-changing operations. Finally, the projection operator $Q$ maps the processed feature back to the target function space, producing the predicted output $G(u)$. Different colors in the diagram represent different block types, visually distinguishing the roles played by each architectural component. The CNO framework thus combines the theoretical foundation of operator learning with the practical strengths of convolutional architectures, enabling scalable and flexible approximations of complex function-to-function mappings.

\subsection{Physics-Informed Loss Formulation}

To enhance the physical consistency and generalization capability of the neural operator, we adopt a physics-informed training strategy that incorporates both data supervision and PDE-based constraints. Specifically, we combine the physics-based loss term \(J_{\text{pde}}\), derived from the governing Helmholtz equation, with the conventional data loss \(J_{\text{data}}\) to train the CNO. This hybrid loss effectively reduces the dependence on large quantities of labeled data, transforming the learning problem from a purely supervised task into a semi-supervised one. It also promotes physically plausible predictions even in data-sparse situations. \par
Although the Physics-Informed CNO (PICNO) can, in principle, be trained solely using the physics constraint \(J_{\text{pde}}\), incorporating the data-driven loss term \(J_{\text{data}}\) provides additional guidance during optimization and introduces for boundary conditions. In practice, the data loss enforces the network toward data-consistent solutions, while the PDE loss enforces physical correctness. The effect of these two components improves convergence stability and enhances the fidelity of the predicted wavefields, especially in the high-frequency regime. More specifically, the data loss term \(J_{\text{data}}\) encourages the model output to align with the ground truth scattered wavefield \(\delta U\), and is defined as the relative mean squared error (MSE) between the network prediction and the reference label:
\begin{equation}
J_{\text{data}} = 
\bigl\| G_{\theta} - \delta U \bigr\|_2^2,
\end{equation}
where \(G_{\theta}\) denotes the scattered wavefield predicted by PICNO, and \(\delta U\) is the reference scattered wavefield obtained from numerical simulation. The physics loss term \(J_{\text{pde}}\) penalizes violations of the governing PDE. It is formulated by using the network output \(G_{\theta}\) into the scattering form of the Helmholtz equation and evaluating the residual over the spatial domain:
\begin{equation}
J_{\text{pde}} = 
\biggl\| \frac{\omega^2}{v^2}\,G_{\theta}
+ \nabla^2 G_{\theta}
+ \omega^2 \left( \frac{1}{v^2} - \frac{1}{v_0^2} \right)\, U_{0} \biggr\|_2^2,
\label{eq8}
\end{equation}
here, \(\nabla^2\) represents the Laplacian operator, and \(U_0\) is the background wavefield generated from the constant background velocity model \(v_0\). The PDE residual is evaluated pointwise over the input grid using an eighth-order finite difference stencil, which provides high numerical accuracy while preserving the spatial resolution of the predicted fields. The total loss used for training is a weighted sum of the two terms:
\begin{equation}
J = J_{\text{data}} + \lambda\, J_{\text{pde}},
\label{eq9}
\end{equation}
where the weight \(\lambda\) balances the contribution of the physics constraint relative to the data supervision. In this study, we determine \(\lambda\) empirically through trial and error. While more advanced hyperparameter optimization strategies may be considered, our empirical tuning yields satisfactory performance. Details regarding the weight selection issue are demonstrated in the "Discussion" section.\par
Overall, by jointly optimizing both data fidelity and physical consistency, the PICNO architecture learns a mapping that not only fits the training data but also adheres to the underlying wave physics. This joint supervision enables the model to generalize better across unseen samples and frequencies, offering a powerful framework for efficient wavefield prediction in complex velocity models.

\section{Results}
In this section, we employ the CNO and its physics-informed variant (PICNO) to learn the solution operator of the frequency-domain Helmholtz equation. In order to increase the training difficulty and maximize the evaluation of PICNO's generalization ability, we design two cases where seismic sources are placed beneath the surface. In Case 1, the horizontal coordinates of the sources are randomly laterally sampled across the model domain, while the depth is fixed at the bottom layer. In Case 2, we consider a more challenging scenario in which the source locations are randomly distributed throughout the entire domain, both horizontally and vertically.

\subsection{Case 1}

\begin{figure}[!t]
    \centering
    \includegraphics[width=1\columnwidth]{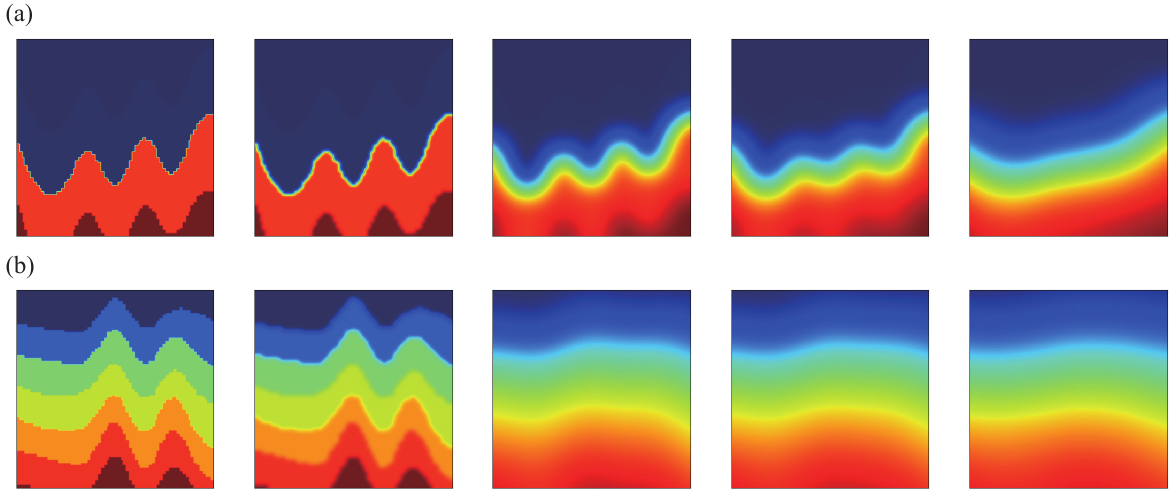}
    \caption{(a) and (b) show two representative velocity models selected from the training dataset, along with their corresponding four smoothed variants. These variations are used for data augmentation to improve the model’s robustness.}
    \label{fig12}
\end{figure}

To account for wavefield variability across frequencies, we consider three discrete frequencies in the Case 1 study: 8 Hz, 10 Hz, and 12 Hz. For each frequency, we construct a relatively small training set covering only 72 velocity models randomly chosen from the OpenFWI dataset \citep{deng2022openfwi}. Furthermore, each velocity model is augmented with four smoothed variants of different smoothing degrees, resulting in five versions per model. In Figure~\ref{fig12}, we show two examples in the training dataset. For each velocity model, we generate ten different source locations, leading to a total of \(72 \times 5 \times 10 = 3600\) training samples per frequency. The validation set for each frequency consists of 16 velocity models, leading to a total of \(16 \times 5 \times 10 = 800\) validation samples per frequency. We separately train the network on the data of these three different frequencies to investigate the impact of frequency on network performance. It is important to note that we deliberately limit the number of base training velocity models to 72 to rigorously evaluate the generalization capability of the network. The addition of smoothed velocity models serves as a data augmentation strategy that helps the network learn the underlying structure of background wavefields, which is particularly beneficial for future inversion tasks \citep{taufik2025wavenumber}. Again, the goal is to train the neural operators to predict the corresponding scattered wavefields \(\delta U\) from the input background wavefield \(U_0\) and the velocity model \(v\). Both CNO and PICNO are optimized using the Adam optimizer, with a fixed learning rate of \(1 \times 10^{-6}\) and a batch size of 32. Each network is trained for 500 epochs per frequency. The training loss is computed as a combination of data misfit and, in the case of PICNO, physics residuals from the PDE constraint.\par

\begin{figure}[!t]
    \centering
    \includegraphics[width=0.90\columnwidth]{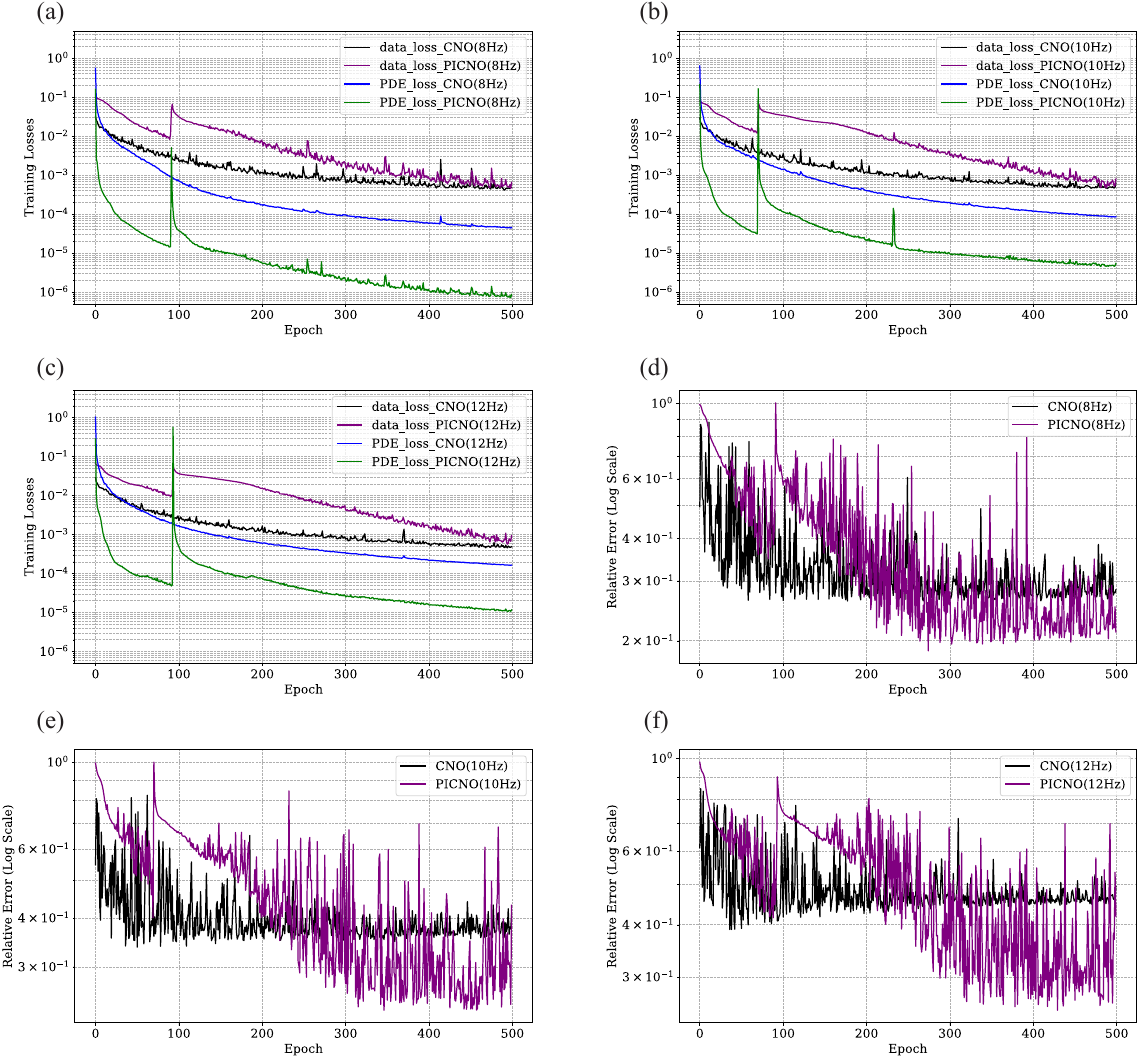}
    \caption{(a)-(c) The training loss curves for the 500 epochs of training for the Helmholtz equation. (d)-(f) Relative $L_{2}$ error curves of CNO and PICNO evaluated on the validation dataset. Here, we compare the different losses from the data and PDE terms for predicting 8Hz, 10Hz, and 12Hz wavefields.}
    \label{fig2}
\end{figure}

We illustrate the training loss curves for the Helmholtz equation for the three frequencies (Figure~\ref{fig2}(a)-(c)). As expected, both CNO and PICNO demonstrate similar performance in minimizing the data loss component, suggesting their capacity to fit the observed training data. However, a key difference emerges in the PDE loss: PICNO consistently achieves lower loss values, indicating that it better satisfies the underlying wave equation. This result suggests that the physics-informed loss not only regularizes the training, but also encourages physically consistent predictions, particularly in velocity models that differ from those seen during training. To further assess generalization performance, we compute the relative \(L_2\) error on the validation set:
\begin{equation}
\text{Relative } L_2 \text{ Error} =\frac{1}{N}\sum_{i=1}^{N} \frac{\| G_{\theta}^{(i)} - \delta U^{(i)} \|_2}{\| \delta U^{(i)} \|_2},
\label{relative_error}
\end{equation}
where \(G_{\theta}^{(i)}\) and \(\delta U^{(i)}\) represent the predicted and true scattered wavefields, respectively, and \(N\) is the number of validation samples. We show the the relative error of these two neural operators in Figure~\ref{fig2}(d)-(f), it is clear that PICNO consistently outperforms CNO in terms of the relative error for three frequencies, demonstrating its superior ability to generalize across different velocity models and frequencies. This improvement highlights the advantage of incorporating physical knowledge into the training process, particularly in low-data regimes. \par

On the other hand, we should also acknowledge some of PICNO’s limitations. Firstly, its validation error tends to exhibit significant instability during training, characterized by abrupt fluctuations. Secondly, the training loss curves occasionally contain outliers or sudden jumps in the loss values. These behaviors are indicative of difficulties in convergence and robustness, particularly in the presence of complex velocity models. A primary contributor to these issues is the physics-informed loss term, which is beneficial for enforcing physical fidelity. However, it could also substantially increase the complexity of the optimization landscape. Unlike purely data-driven loss functions that often produce relatively smooth gradients, the PDE residual terms may introduce highly non-convex structures and sharp local minima. As a result, the optimization algorithm—such as Adam—may struggle to find a stable descent path, especially when the PDE loss is weighted significantly. Moreover, the PDE loss introduces constraints that are sensitive to both frequency ($\omega$ in Equation~\ref{eq8}) and complex structure in the velocity models ($v$ in Equation~\ref{eq8}). This phenomenon can create steep gradients, making the learning dynamics highly sensitive to initialization, learning rate, and the balance between data and physics terms ($\lambda$ in Equation~\ref{eq9}). \par

To mitigate these challenges, more advanced neural operator frameworks may be required. Alternatively, preconditioning techniques or curriculum learning schemes could help ease the optimization process by gradually increasing the contribution of the PDE loss \citep{cheng2025seismic}. Overall, while PICNO demonstrates strong potential in incorporating physical constraints, its training behavior highlights the need for more robust and scalable strategies to navigate the intricate loss surface introduced by physics-informed learning. \par

\begin{figure}[!t]
    \centering
    \includegraphics[width=1\columnwidth]{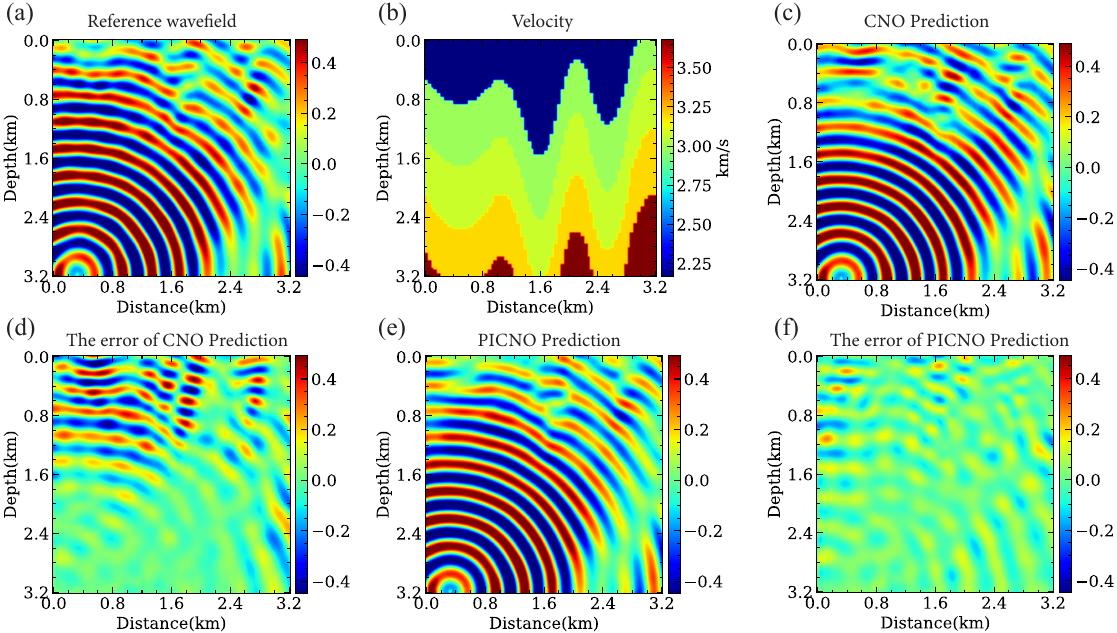}
    \caption{Prediction results of frequency-domain wavefields for a test sample at 8 Hz. (a) Reference wavefield computed using a conventional numerical solver; (b) corresponding velocity model; (c) wavefield predicted by CNO; (d) difference between the CNO prediction and the reference wavefield; (e) wavefield predicted by PICNO; and (f) difference between the PICNO prediction and the reference wavefield. This ordering is consistent across all wavefield prediction figures presented in this work.}
    \label{fig4}
\end{figure}

\begin{figure}[!t]
    \centering
    \includegraphics[width=1\columnwidth]{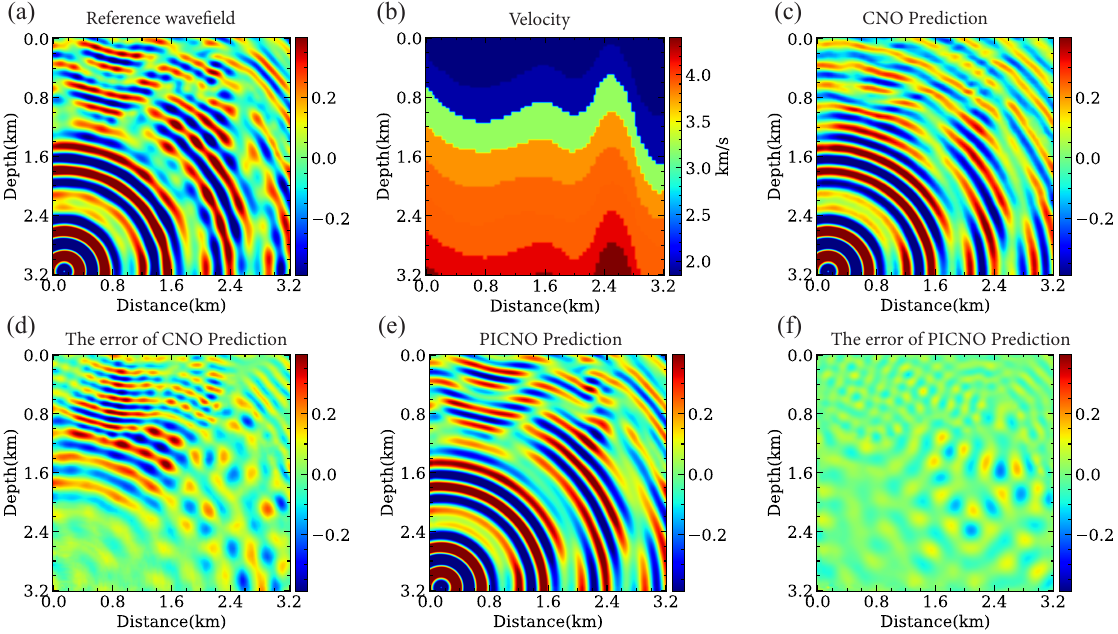}
    \caption{The frequency domain wavefield prediction results for an instance from the test set (10Hz).}
    \label{fig5}
\end{figure}

\begin{figure}[!t]
    \centering
    \includegraphics[width=1\columnwidth]{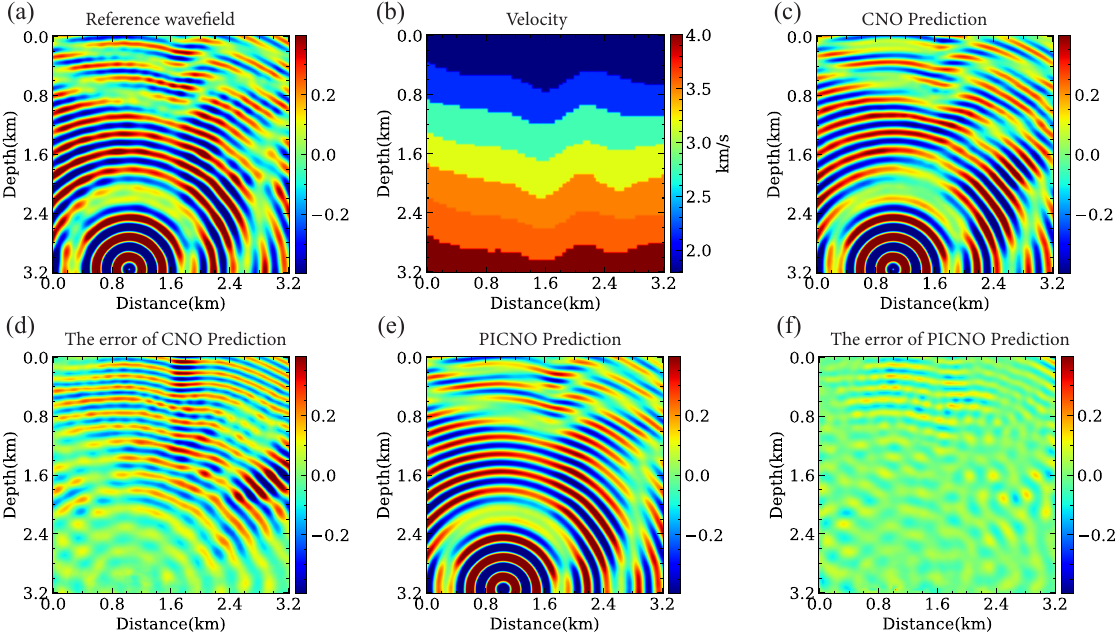}
    \caption{The frequency domain wavefield prediction results for an instance from the test set (12Hz).}
    \label{fig6}
\end{figure}

We further evaluate the performance of both networks using 800 test samples (velocity models not included in either the training or validation sets). Table~\ref{tab1} presents the average prediction error and the relative improvement achieved by PICNO over CNO. Additionally, Figures~\ref{fig4}–\ref{fig6} illustrate the predicted wavefields at three different frequencies—8~Hz, 10~Hz, and 12~Hz—on three representative unseen velocity models. We emphasize that both neural operators require only approximately 0.83 seconds to produce the frequency-domain wavefield solution, whereas the conventional numerical method takes around 3 seconds. This highlights another key advantage of neural operators—their ability to rapidly generate solutions to partial differential equations. From these figures, it is evident that PICNO consistently outperforms CNO across all frequencies. At \textbf{8~Hz}, both models provide visually reasonable approximations of the reference wavefield, but PICNO shows noticeably smaller residuals, particularly in the shallow regions, where the wavefield exhibits more complex waveform patterns (Figure~\ref{fig4}(e)). At \textbf{10~Hz}, the difference between the two models becomes more apparent: CNO begins to exhibit visible phase mismatches and amplitude damping (Figure~\ref{fig5}(c)), while PICNO maintains both waveform continuity and high-frequency detail (Figure~\ref{fig5}(e)). The superiority of PICNO is most obvious at \textbf{12~Hz}, where high-frequency components dominate the wavefield. In this case, CNO struggles to recover small-scale features, resulting in significant structural deviations (Figure~\ref{fig6}(c)). In contrast, PICNO successfully captures wavefront details and complex diffraction patterns, demonstrating its enhanced generalization capacity and ability to honor physical constraints (Figure~\ref{fig6}(e)). These qualitative observations, supported by the quantitative statistics in Table~\ref{tab1}, affirm that PICNO not only reduces prediction errors but also yields more physically consistent and high-resolution wavefield predictions, particularly under high-frequency conditions where accurate modeling becomes increasingly challenging.

\begin{table}[H]
    \centering
    \resizebox{0.8\columnwidth}{!}{%
    \begin{tabular}{|l|c|c|c|}
        \hline
        {MODEL} & {8 Hz} & {10 Hz} & {12 Hz} \\
        \hline
        CNO  & $0.25$ & $0.34$ & $0.50$ \\
        \textbf{PICNO} & $\mathbf{0.18}$ & $\mathbf{0.22}$ & $\mathbf{0.23}$ \\
        \hline
        \text{Relative Promotion } & {$27.4\%$} & {$35.2\%$} & {$53.1\%$} \\
        \hline
    \end{tabular}
    }
    \caption{The overall relative \(L_2\) error (Equation~\ref{relative_error}) of CNO and PICNO on test samples. Promotion refers to the relative error reduction. }
    \label{tab1}
\end{table}

\subsection{Case 2}

\begin{figure}[!t]
    \centering
    \includegraphics[width=1\columnwidth]{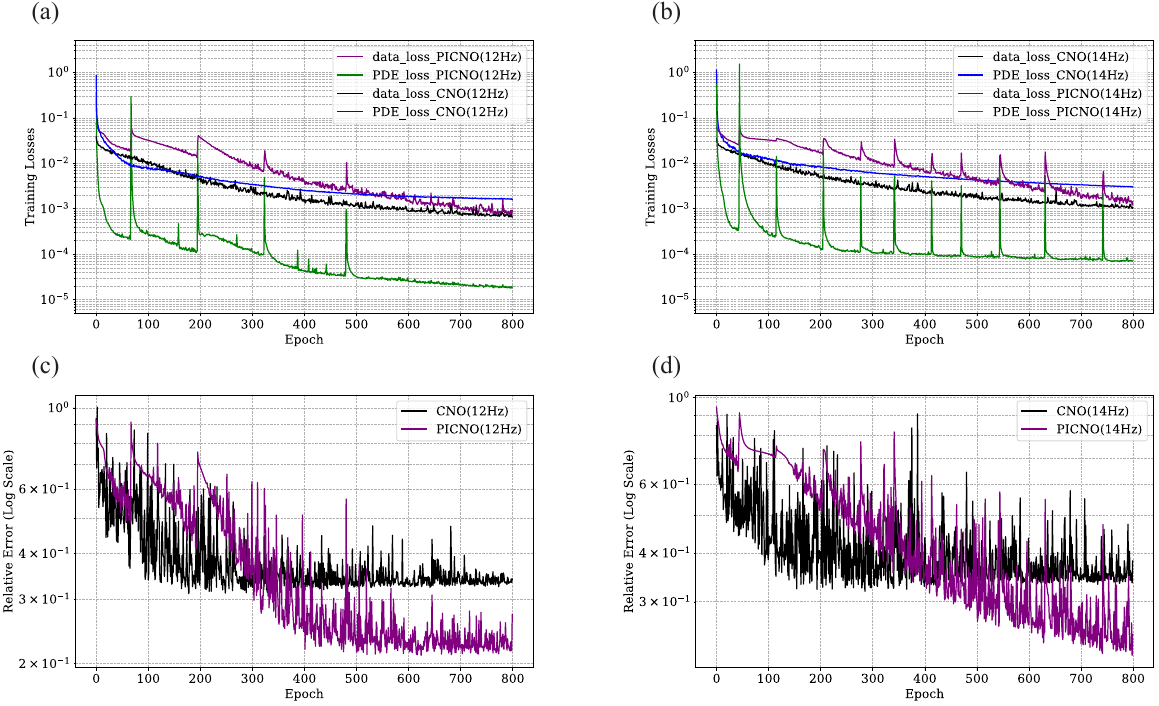}
    \caption{(a) and (b) show the training loss curves for the 800 epochs of training for the Helmholtz equation. (c) and (d) show Relative L2 error curves of CNO and PICNO evaluated on the validation dataset. Here, we compare the different losses for predicting 12Hz and 14Hz wavefields.}
    \label{fig7}
\end{figure}

To further assess the generalization capability of PICNO, we design another experiment in which source locations are randomly distributed throughout the entire spatial domain. In this setting, the source positions vary in both the depth and lateral dimensions across different velocity models. We generate 3200 frequency-domain wavefields using 72 velocity models for each frequency as stated in Case 1 and retrain the PICNO model. Each validation set consists of 800 wavefields generated using 16 velocity models, where the source locations also vary along the depth and lateral directions. During the training stage, we keep all hyperparameters identical to those used in the previous section, except that the number of training epochs is set to 800. This adjustment is made because we observe that the neural operators tend to converge more slowly when source locations are randomly distributed within the full model space. Figure~\ref{fig7} shows the training loss and validation mean square error for 12 and 14 Hz wavefields. It is evident that, despite being trained on a relatively limited dataset, PICNO achieves considerably lower PDE loss than CNO, as illustrated in Figure~\ref{fig7}(a) and Figure~\ref{fig7}(b). This demonstrates that PICNO once again captures the physical principles encoded in the PDE loss. This observation is further supported by the comparison of validation relative errors in Figure~\ref{fig7}(c) and Figure~\ref{fig7}(d), where PICNO achieves substantially lower validation errors than CNO. \par

\begin{figure}[!t]
    \centering
    \includegraphics[width=0.95\columnwidth]{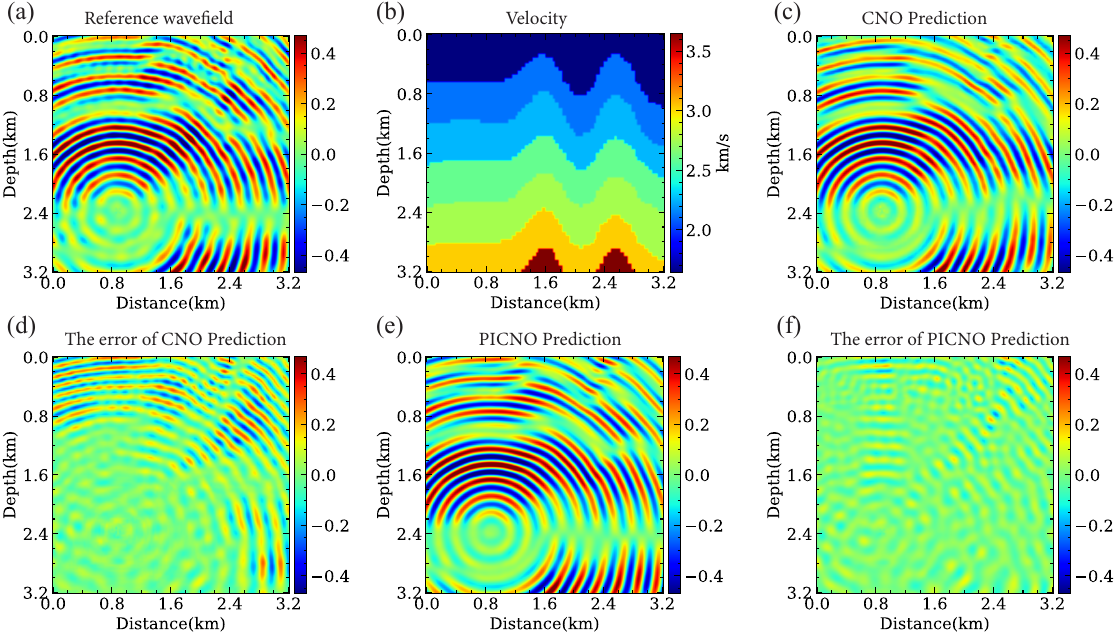}
    \caption{The frequency domain wavefield prediction results for an instance from the test set (12Hz).}
    \label{fig8}
\end{figure}

\begin{figure}[!t]
    \centering
    \includegraphics[width=0.95\columnwidth]{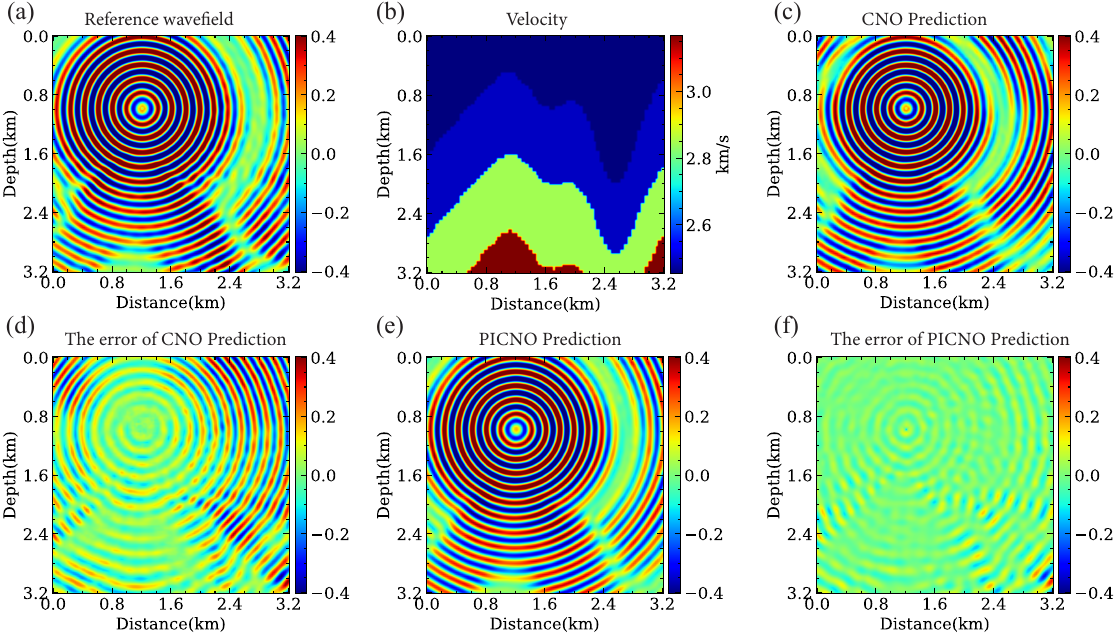}
    \caption{The frequency domain wavefield prediction results for an instance from the test set (14Hz).}
    \label{fig9}
\end{figure}

We also randomly select velocity models from the test dataset for evaluation. As shown in Figure~\ref{fig8} and Figure~\ref{fig9}, PICNO provides more accurate wavefield predictions compared to CNO. In particular, PICNO captures wavefield details that are closer to the reference wavefield simulated by solving the physical equations. More precisely, the predictions made by CNO deviate from the expected physical behavior, with the predicted wavelengths appearing more random. In contrast, PICNO maintains high fidelity across the domain, preserving both the amplitude and phase structures even under high-frequency, random source location conditions. This indicates that the inclusion of physics constraints not only regularizes the training process but also significantly improves the model's generalization ability. It enables PICNO to handle challenging wave propagation settings where the source locations vary. Thus, PICNO proves to be a more reliable and physically faithful surrogate for wavefield simulation in practical geophysical applications.

\subsection{Out of distribution test}

\begin{figure}[!t]
    \centering
    \includegraphics[width=1\columnwidth]{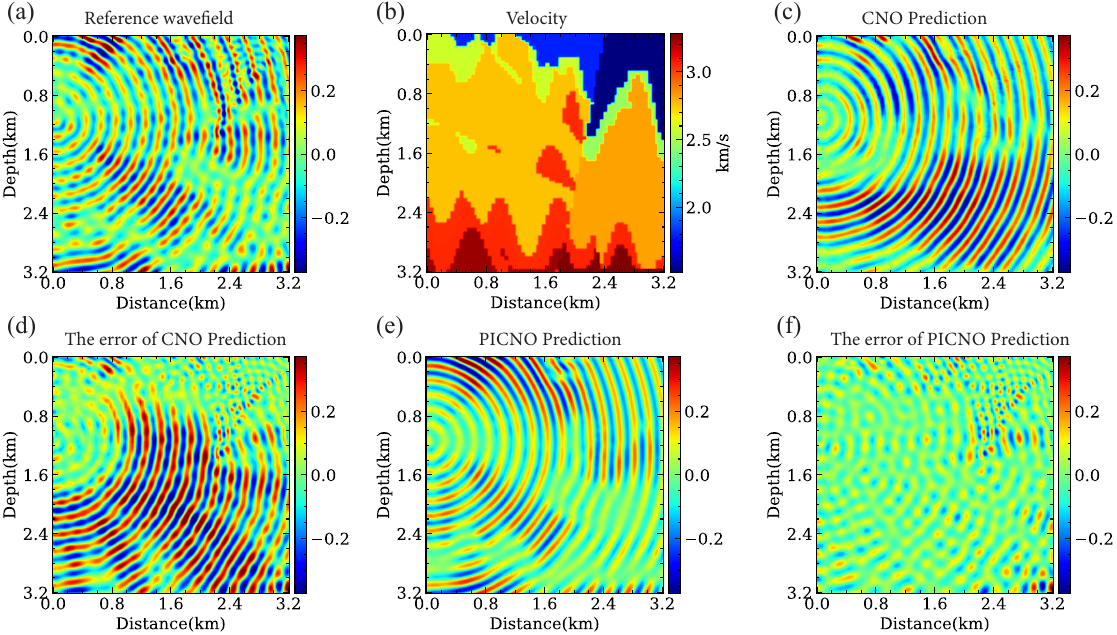}
    \caption{The frequency domain wavefield prediction results for an OOD instance from the test set (14Hz).}
    \label{fig10}
\end{figure}

In addition to the in-distribution velocity model tests, we further assess the generalization capability of PICNO through an out-of-distribution (OOD) experiment. The velocity model used in this evaluation, as shown in Figure~\ref{fig10}(b), incorporates geological structures such as faults and sharp velocity discontinuities—features that are not included in the training dataset. Despite the presence of such unseen and complex geophysical patterns, PICNO remains robust and consistently yields wavefield predictions that closely match the ground truth obtained from numerical solutions of the Helmholtz equation (Figure~\ref{fig10}(e)). In contrast, CNO exhibits substantial prediction errors (Figure~\ref{fig10}(c)). The improved performance of PICNO in these OOD scenarios highlights its ability to internalize the governing physics through the incorporation of PDE-based constraints. Compared with Figure~\ref{fig10}(d) and Figure~\ref{fig10}(f), although PICNO is trained on a comparatively small set of velocity models, it demonstrates a clear capacity to generalize across diverse velocity models. However, we should also note that the wavefield predictions produced by PICNO appear smoother compared to the reference wavefields. This is likely due to the limited diversity of the training data as well as the inherent smoothing effect of the convolution operations employed in the CNO architecture. Nevertheless, PICNO is still well-suited for real-world geophysical applications where the subsurface structures are often unknown. The improved performance of PICNO in these OOD scenarios highlights its ability to learn the governing physics through the incorporation of PDE-based constraints during training. Compared with CNO, which relies solely on data-driven supervision, PICNO benefits from an inductive bias toward physically consistent solutions. Notably, despite being trained on a relatively small and geologically homogeneous set of velocity models, PICNO exhibits a clear capacity to generalize across structurally diverse models with minimal degradation in accuracy.

\section{Discussion}
In the “Example” section, we have demonstrated the superior performance of PICNO in predicting frequency-domain wavefields. However, despite its advantages, PICNO also introduces several challenges. In this section, we focus on analyzing and discussing critical limitations observed during training, including the sensitivity of the model to PDE loss weight configurations.

\subsection{The balance between the PDE and data losses}

\begin{figure}[!t]
    \centering
    \includegraphics[width=0.6\columnwidth]{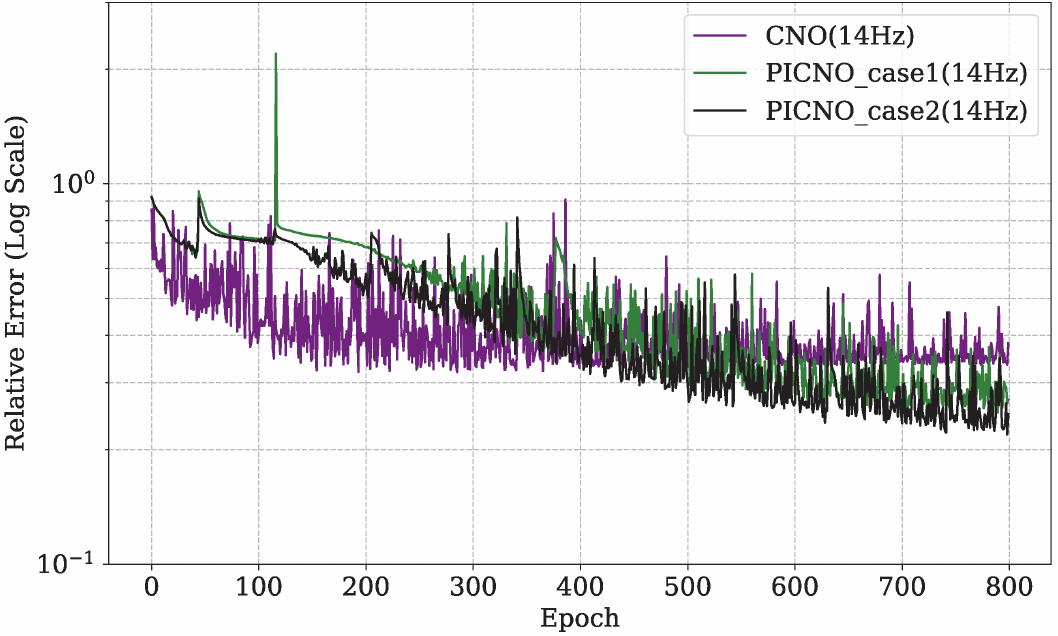}
    \caption{ Validation relative error curves under three different training configurations. The results compare CNO (trained with data loss only), PICNO\_case1 (trained with a PDE-to-data loss ratio of 10:1), and PICNO\_case2 (with a ratio of 8:1). This figure illustrates how the choice of loss weighting influences convergence behavior and generalization performance during training.}
    \label{fig13}
\end{figure}

During training, we observe that the relative weighting between the data loss and the physics-informed loss plays a crucial role in determining both the convergence behavior and the inference performance of the neural operator. In particular, putting an excessively large or small weight on the PDE loss term can lead to either underfitting the physical constraints or a degradation in data fidelity. This delicate balance is illustrated in Figure~\ref{fig13}, where we compare the validation relative error across three configurations: "CNO": solely with a data loss term; "PICNO\_case1": with a PDE-to-data loss ratio of 10:1; and "PICNO\_case2": with a ratio of 8:1. While "PICNO\_case2" achieves slightly lower validation error, the variation in performance highlights the sensitivity of the network to the weighting of the physics term. It is important to emphasize that no universally optimal ratio generalizes across different datasets or frequencies. Instead, the selection of this hyperparameter often relies on empirical tuning and trial-and-error. It is also recognized that the process of tuning loss weights through empirical strategies may lead to significant computational costs in some real-world applications. The solution to this problem remains an open question and requires further investigation.\par

\subsection{Outlook on Physics-Informed Neural Operators in Geophysics}

The application of physics-informed neural operators (PINOs) in seismology and geophysical problems remains in its early stages \citep{okazaki2024scientific}. several open questions must be addressed to enable broader application. One major challenge lies in formulating appropriate partial differential equations that accurately capture the underlying physics of complex Earth structures \citep{alkhalifah1995velocity, alkhalifah2000acoustic}. Also, the limited availability of labeled data remains a significant constraint on the applicability of neural operators in geophysical and seismological problems. Another key issue is the scalability of these models, both in terms of computational efficiency and training robustness. Nevertheless, embedding physical laws directly into neural network architectures offers a compelling paradigm for scientific machine learning. By constraining learning with domain knowledge, PINOs have the potential to generalize beyond the training data and provide physically consistent predictions, even in out-of-distribution scenarios. In the future, we aim to further explore the potential of physics-informed neural operators for the geophysical inversion problem.

\section{Conclusions}
We introduced a novel physics-informed convolutional neural operator to learn the Helmholtz equation efficiently. Our approach not only relies on the neural operator to learn data fitting but also allows it to capture the underlying physical laws, significantly enhancing its generalization ability. Our experiments demonstrate that even with a limited amount of training data, PICNO achieves remarkable improvements in wavefield prediction compared to the standard convolutional neural operator. The advantage is particularly evident in high-frequency scenarios, where traditional neural operators often struggle due to increased complexity. The ability to generalize across different complex velocity models and source locations, even at relatively high frequencies, makes it highly promising for applications in seismic inversion.

\section*{Open Research Section}
The code supporting the findings of this study is available at https://github.com/DeepWave-KAUST/PICNO-pub.

\acknowledgments
The authors thank KAUST for providing the computational resources and research support that made this work possible.

%
%

\bibliography{agusample}  

%
%

%
%

\end{document}